\documentclass{ws-rv961x669}
\usepackage{ws-rv-van}     
\usepackage{ws-rv-thm}     
\usepackage{subfigure}     
\makeindex
\begin{document}
\vskip2cm
\chapter{The Physics Case for the $\sqrt{s_{NN}} \approx 10$ GeV Energy Region.}
%
\author{J.\ Cleymans}
\address{UCT-CERN Research Centre and Physics Department, \\
University of Cape Town,\\
Rondebosch 7701, South Africa}
\begin{center}
{\sl To the memory of Professor Dr. Walter Greiner}
\end{center}
\begin{abstract}
There  are  indications that the beam energy region $\sqrt{s_{NN}} \approx 10$ GeV  
for heavy-ion collisions is an   interesting one.
The final state has the highest net baryon density at this beam energy. 
A transition from a baryon dominated to a meson dominated final state
takes place around this beam energy. 
Ratios of strange particles to mesons show clear and pronounced maxima
around this beam energy.
The theoretical interpretation can be clarified by covering fully this 
energy region.
In particular the strangeness content needs to be determined, data covering
the full phase space ($4 \pi$) would be helpful to establish the 
properties of this energy region. 
\end{abstract}

\body
\section{Introduction}
Heavy-ion collisions~\cite{Stoecker:1986ci}  at high energies produce a large numbers of secondaries. At the LHC the 
number of charged particles produced in Pb-Pb collisions at 5.02 TeV~\cite{dNch} is shown in Fig.~1,
thus, including neutral particles, a total of approximately 30 000 particles is being produced on average in such a collision.
\begin{figure}[htb]
\centerline{\includegraphics[height=8cm,width=8cm]{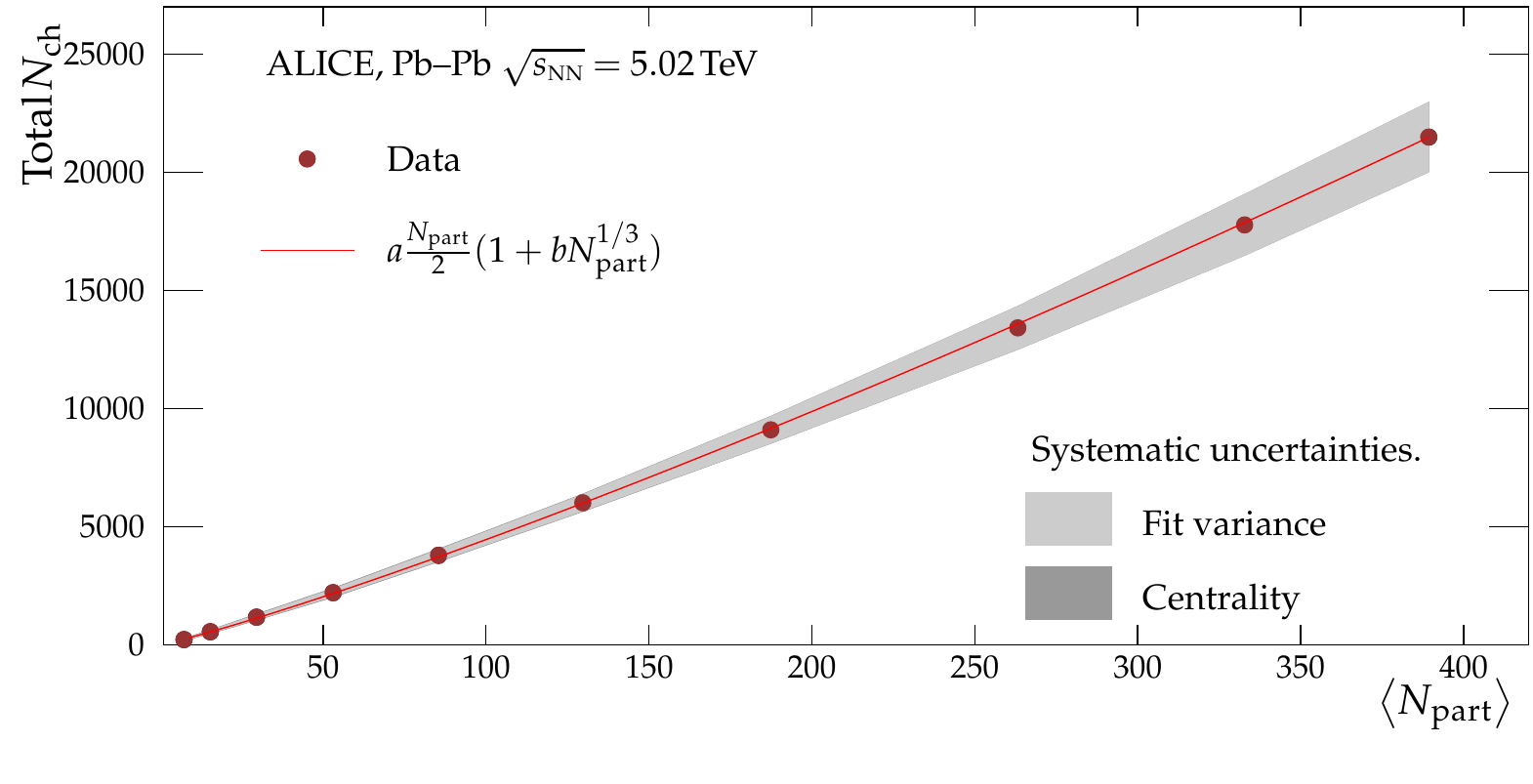}}
\caption{Number of charged particles produced in a Pb-Pb collision as a function of beam energy as measured by the 
ALICE collaboration~\cite{dNch} .}
\label{e_1999}
\end{figure}
It is  natural to try a statistical-thermal model to analyze these. 
As it turns out such an analysis is useful for a very wide range of beam energies, stretching from 
 1 GeV  all the way up  to the highest energies available at the LHC.
For such an analysis one has to keep in mind that
a relativistic heavy-ion collision passes through several stages.
At one of the later, hadronic, stages, the system is assumed to be dominated by hadronic resonances, on which  the thermal model 
focuses.
The identifying feature of the thermal model  is that all the resonances as listed in~\cite{PDG}  are
assumed to be in thermal and chemical equilibrium.
This  assumption drastically reduces the number of free parameters and thus this stage is determined by just a few
thermodynamic variables namely, the chemical freeze-out temperature $T$, the various chemical potentials $\mu$ determined by
the conserved quantum numbers and by the volume $V$ of the system.
It has been shown that this description is also the correct
one~\cite{Cleymans:1999st,Broniowski:2001we,Akkelin:2001wv} for a scaling expansion as first discussed by
Bjorken~\cite{Bjorken:1982qr}.

In  relativistic heavy ion collisions a new dimension was given to the model
by the highly successful analysis of particle yields, leading to the notion of chemical 
equilibrium which is now a  well-established one in the analysis of relativistic heavy ion 
collisions, see e.g.~\cite{Adamczyk:2017iwn,floris,HADES}. 
\begin{figure}
\centerline{\includegraphics[width=12cm,height=12cm]{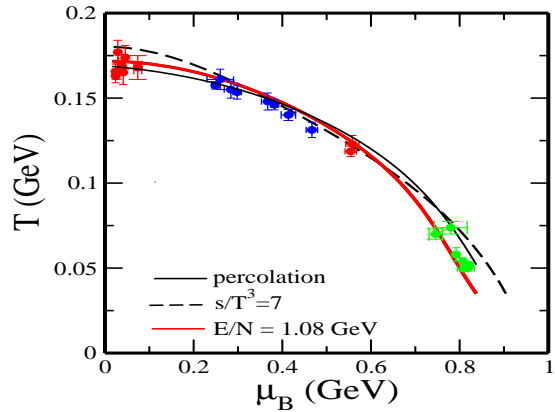}}
\vspace{-4cm}
\caption{Chemical freeze-out temperature $T$ vs. the baryon chemical potential  at different beam 
energies together with curves corresponding to a fixed ratio of energy per hadron divided by
total number of hadrons in the resonance gas before decay of resonances~\cite{wheaton} .
Also shown are calculations based on the percolation model~\cite{Magas:2003wi} and for a fixed value of the entropy density 
divided by $T^3$.}
\label{criteria}
\end{figure}
In view of the success of  chemical freeze-out  in relativistic heavy ion collisions, 
much effort has gone into finding models that describe this chemical 
freeze-out, a comparison~\cite{wheaton} of three parameterizations is shown in Fig.~\ref{criteria}.

There are of course uncertainties in the thermal model, 
one of these  is about the decays of resonances, another one is whether some resonances exist or not~\cite{PDG} .
Particle yields are determined from:
$$
N_i  = \sum_j N_j Br( j \rightarrow i)  .
$$
Hence a lack of knowledge of branching ratios  affects the quality of results obtained from the thermal model.\\[0.2cm]
As an example, the final yield of $\pi^+$'s is given by
$$
N_{\pi^+}= N_{\pi^+}({\mathrm{thermal}})+N_{\pi^+}({\mathrm{resonance~decays}})
$$
and, depending on the temperature, over 80\% of observed pions could be due to resonance decays. 
Hence the crucial importance of these decays.
Various theoretical uncertainties have been recently discussed in~\cite{alba} .
\section{What makes  the beam energy $\sqrt{s_{NN}} \approx 10 $ GeV special?}
\subsection{Maximum net baryon density}
\begin{figure}[htb]
\vspace{-2cm}
\includegraphics[width=\linewidth,height=12cm]{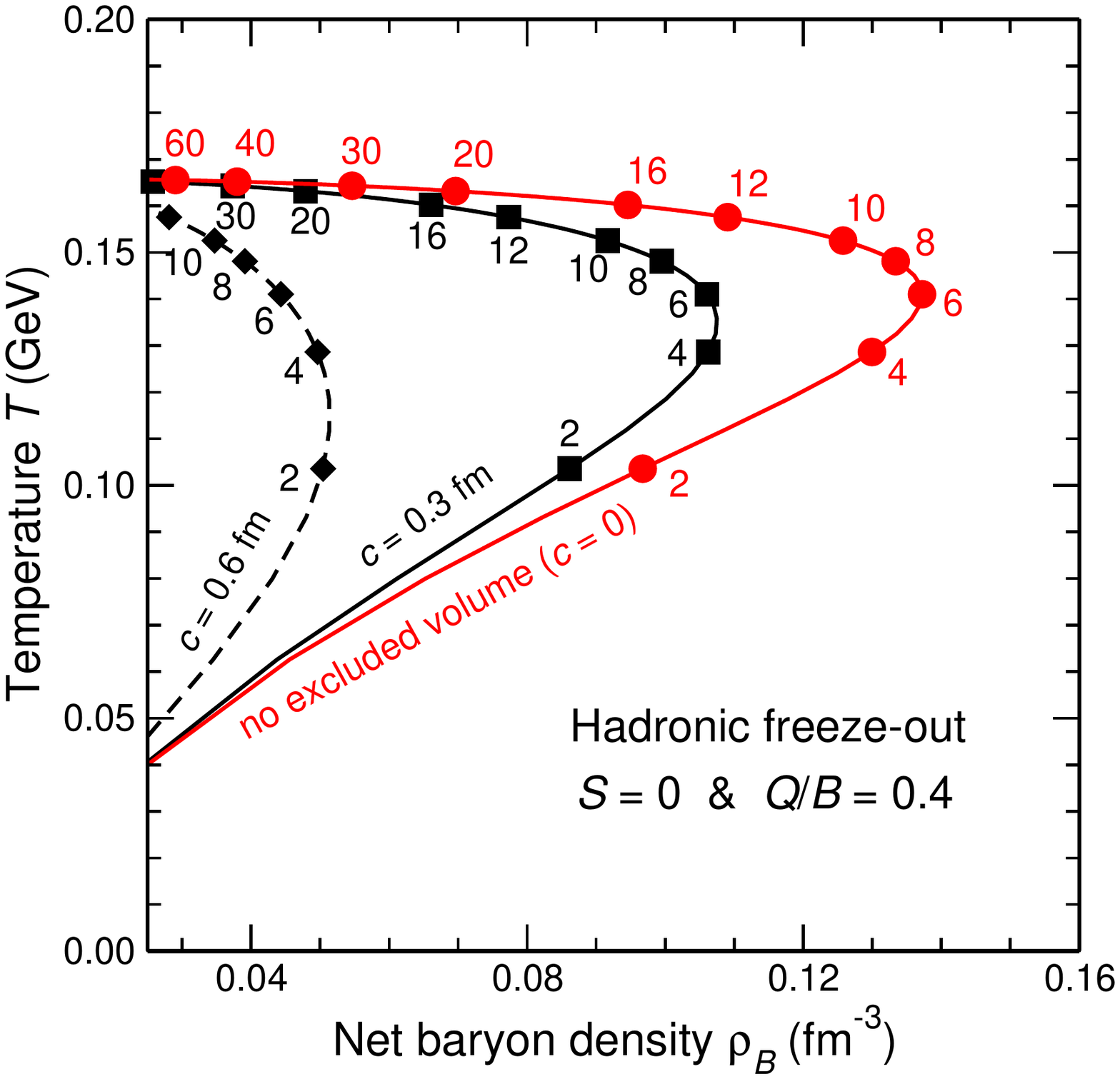}
\caption{The hadronic freeze-out line in the $\rho_B-T$ phase plane 
as obtained from the values of $\mu_B$ and $T$
 that have been extracted from the experimental data in~\cite{wheaton} .
The calculation employs values of $\mu_Q$ and $\mu_S$ 
that ensure $\langle S\rangle=0$ and $\langle Q\rangle=0.4\langle B\rangle$
for each value of $\mu_B$~\cite{randrup2}.  
Also indicated are the beam energies 
for which the particular freeze-out 
conditions are expected. The dependence on a hard-core radius is indicated. 
}
\label{randrup_figure}
\end{figure}

The resulting freeze-out curve  in the $T-\mu_B$ plane, shown in Fig.~\ref{criteria}, can also be drawn in the
temperature $T$ vs net baryon density plane as was done in~\cite{randrup2} . The
resulting curve is shown in Fig.~\ref{randrup_figure}.  At very high beam energies the net baryon density is zero  
because equal numbers of particles and antiparticles are being produced while
at low temperatures the net baryon density is very high. Fig.~\ref{randrup_figure} shows that the a clear maximum
exists just below the $\sqrt{s_{NN}} = 10 $ GeV beam energy region.
\subsection{Transition from a baryon dominated to a meson dominated final state}
\begin{figure}
\centerline{\includegraphics[width=10cm,height=10cm]{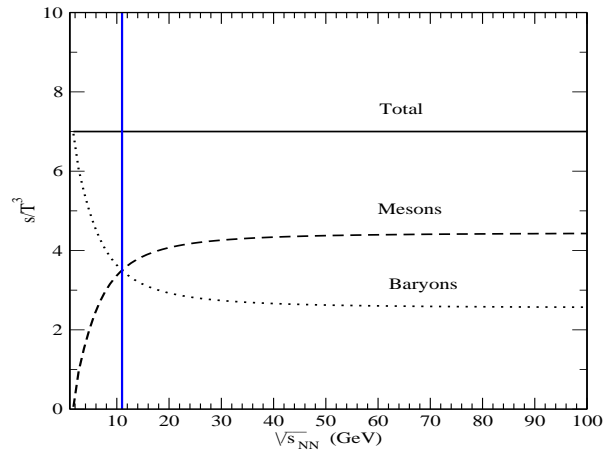}}
\caption{The $s/T^3$ ratio calculated in the thermal model along the constant value consistent with
chemical freeze-out. Also shown are the contributions from the mesons and the baryons.}
\label{sovert3}
\end{figure}

A fairly good criterium for chemical  freeze-outis the constant value of the entropy density
divided  $s/T^3 = 7$ ratio as can be seen from Fig.~\ref{criteria} .
The components that make up the entropy density are shown in Fig.~\ref{sovert3},  the change from a baryon-dominated to a meson-dominated
final state also happens around a beam energy of $\sqrt{s_{NN}} \approx 10 $ GeV.
\subsection{Ratios of strange hadrons to pions}
\begin{figure}
\begin{minipage}[b]{0.45\linewidth}
\centering
  \includegraphics[width=\textwidth,height=0.3\textheight]{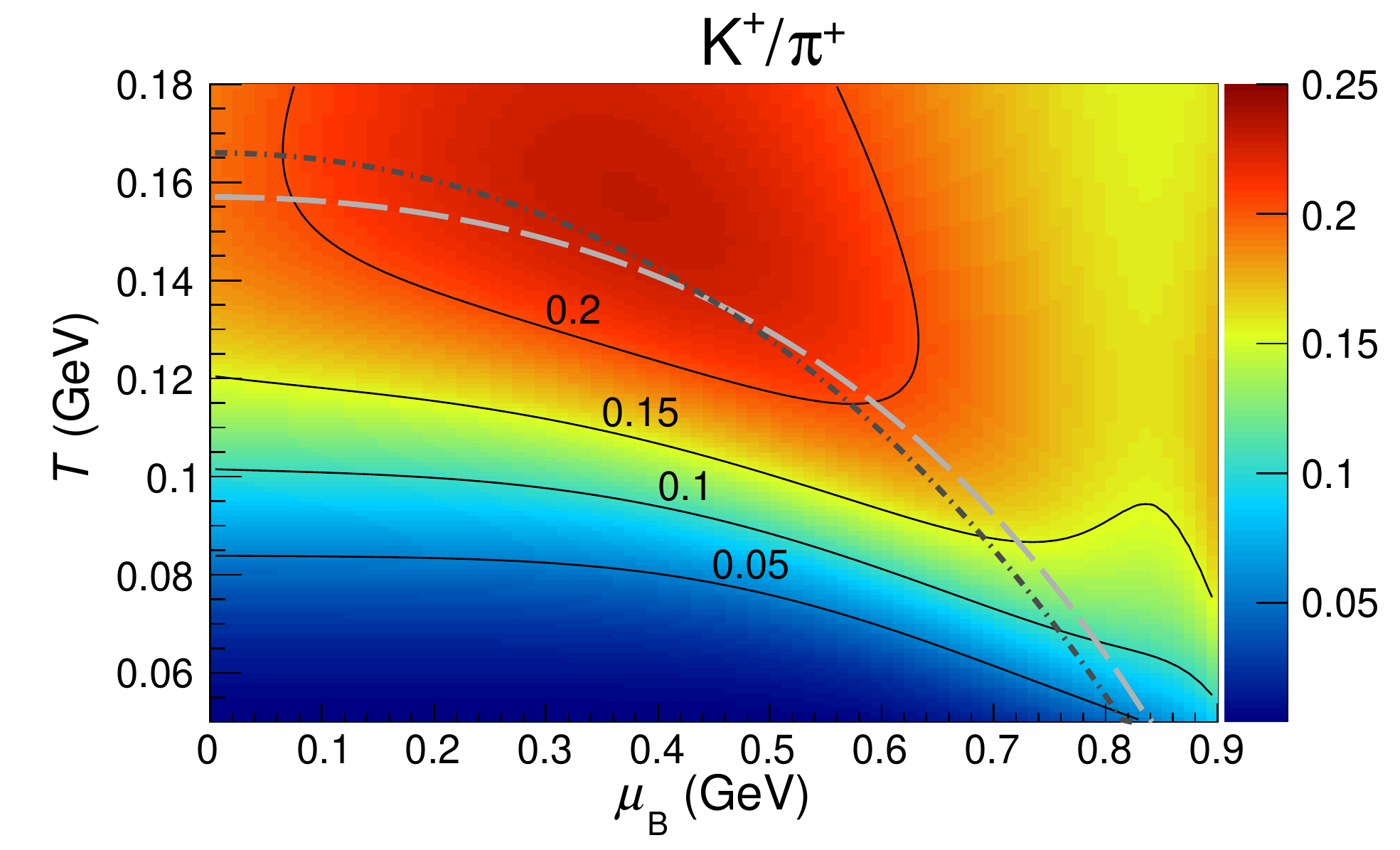}
\end{minipage}
\hspace{0.5cm}
\begin{minipage}[b]{0.45\textwidth}
\centering
   \includegraphics[width=\textwidth,height=0.3\textheight]{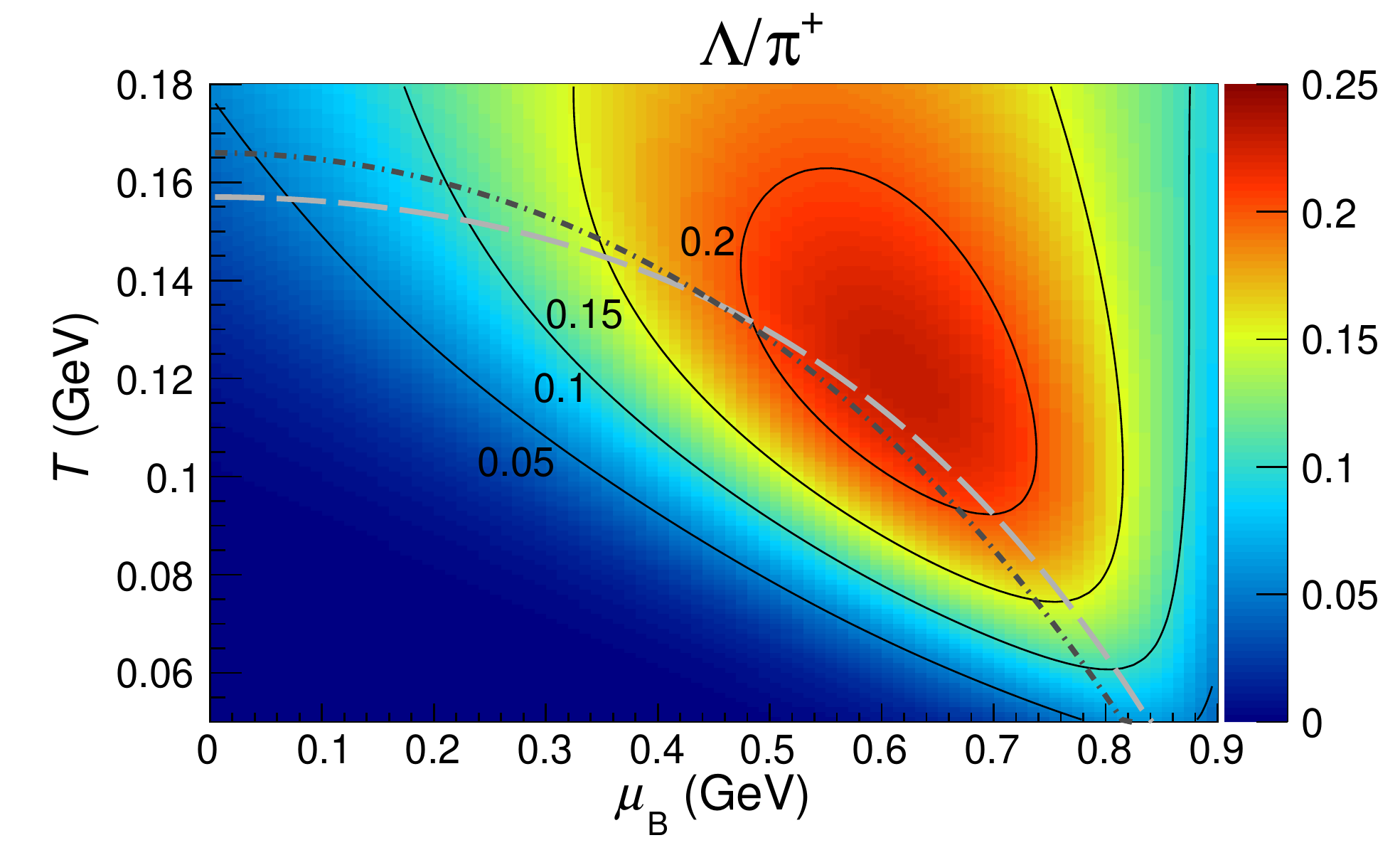}
\end{minipage}
\caption{Lines of constant values of the $K^+/\pi^+$ (left panel) and the $\Lambda/\pi^+$ (right panel) ratios in the $T-\mu_B$ 
plane showing a clear maximum 
in each ratio close to the boundary given by the chemical freeze-out line but in a different position~\cite{Oeschler:2017bwk} .}
\label{kpluspiplus}
\end{figure}

Despite the smoothness in the thermal freeze-out parameters as a function of beam energy, 
strong changes are observed in several  particle ratios, e.g. the horn in the $K^+/\pi^+$ ratio and a similar
strong variation in the $\Lambda/\pi$ ratio~\cite{NA49} .
These  are not observed in $p-p$ collisions, in Pb-Pb collisions they happen at a beam energy of around
$\sqrt{s_{NN}} \approx 10$ GeV.
Within the framework of thermal models this variation has been connected to a change from
a baryon dominated to a meson dominated hadron gas~\cite{transition} . 
The values of the $K^+/\pi^+$ and $\Lambda/\pi^+$ ratios~\cite{Oeschler:2017bwk} 
are shown in Fig.~\ref{kpluspiplus}.
From the lines of constant values for these ratios 
it can be seen that the 
maxima in the thermal model hug the chemical freeze-out line.  It is also important to note that the maxima occur
 for different values of $T$ and $\mu_B$.
\section{Conclusions}
In the thermal model
a change is expected as the hadronic gas undergoes a
transition from a baryon-dominated  to a meson-dominated gas. 
The strong variations seen in the particle ratios
coincide with this transition.
This transition occurs at a
\begin{itemize}
\item temperature $T = $ 151 MeV, 
\item baryon chemical potential $\mu_B = $ 327 MeV, 
\item  energy $\sqrt{s_{NN}} = $ 11 GeV. 
\end{itemize}
There  are thus several   indications that the energy region around 10 GeV, 
covered by  proposed new facilities, is an extremely 
interesting one.
The theoretical interpretation can be clarified by covering this 
energy region.
In particular the strangeness content needs to be determined, data covering
the full phase space (4$\pi$) would be very helpful to determine the thermal parameters of 
a possible phase transition and the existence of a quarkyonic phase as has been discussed recently in~\cite{mclerran} .
\newpage
\bibliographystyle{ws-rv-van}
\bibliography{cleymans_frankfurt_2017_bibtex}

\begin{thebibliography}{10}
\expandafter\ifx\csname urlstyle\endcsname\relax
  \providecommand{\doi}[1]{doi:\discretionary{}{}{}#1}\else
  \providecommand{\doi}{doi:\discretionary{}{}{}\begingroup
  \urlstyle{rm}\Url}\fi

\bibitem{Stoecker:1986ci}
H.~Stoecker and W.~Greiner, {\em Phys. Rept.} {\bf 137}, 277  (1986).

\bibitem{dNch}
ALICE Collaboration, J.~Adam {\em et~al.}, {\em Phys. Lett. B} {\bf 772},   567
   (2017).

\bibitem{PDG}
Particle Data Group Collaboration, C.~Patrignani {\em et~al.}, {\em Chin.
  Phys.} {\bf C40},   100001  (2016).

\bibitem{Cleymans:1999st}
J.~Cleymans and K.~Redlich, {\em Phys. Rev.} {\bf C60},   054908  (1999).

\bibitem{Broniowski:2001we}
W.~Broniowski and W.~Florkowski, {\em Phys. Rev. Lett.} {\bf 87},   272302
  (2001).

\bibitem{Akkelin:2001wv}
S.~Akkelin, P.~Braun-Munzinger and Y.~M. Sinyukov, {\em Nucl. Phys. A} {\bf
  710},   439  (2002).

\bibitem{Bjorken:1982qr}
J.~Bjorken, {\em Phys. Rev. D} {\bf 27},   140  (1983).

\bibitem{Adamczyk:2017iwn}
STAR Collaboration, L.~Adamczyk {\em et~al.}, {\em Phys. Rev.} {\bf C96},
  044904  (2017).

\bibitem{floris}
ALICE Collaboration, M.~Floris {\em et~al.}, {\it Quark matter 2014}, in {\em
  The 24th International Conference on Ultrarelativistic Nucleus-Nucleus
  Collisions.\/},  (Darmstadt, Germany, 2014).
\newblock pp. c103--c112.

\bibitem{HADES}
HADES Collaboration, M.~M.~Lorenz {\em et~al.}, {\it Sqm2015}, in {\em 15th
  International Conference on Strangeness in Quark Matter.\/},  (Dubna, Russia,
  2015).
\newblock p. 012022.

\bibitem{wheaton}
J.~Cleymans, H.~Oeschler, K.~Redlich and S.~Wheaton, {\em Phys. Rev. C} {\bf
  73},   034905  (2006).

\bibitem{Magas:2003wi}
V.~Magas and H.~Satz, {\em Eur. Phys. J.} {\bf C32}, 115  (2003).

\bibitem{alba}
L.~M. Satarov, V.~Vovchenko, P.~Alba, M.~I. Gorenstein and H.~Stoecker, {\em
  Phys. Rev.} {\bf C95},   024902  (2017).

\bibitem{randrup2}
J.~Randrup and J.~Cleymans, {\em Eur. Phys. J. C} {\bf 52},   218  (2016).

\bibitem{NA49}
NA49 Collaboration, C.~Alt {\em et~al.}, {\em Phys. Rev. C} {\bf 77},   024903
  (2008).

\bibitem{transition}
J.Cleymans, H.~Oeschler, K.~Redlich and S.~Wheaton, {\em Phys. Lett. B} {\bf
  615},  ~50  (2005).

\bibitem{Oeschler:2017bwk}
H.~Oeschler, J.~Cleymans, B.~Hippolyte, K.~Redlich and N.~Sharma, {\em Eur.
  Phys. J.} {\bf C77},   584  (2017).

\bibitem{mclerran}
A.~Andronic {\em et~al.}, {\em Nucl. Phys. A} {\bf 837},  ~65  (2010).

\end{thebibliography}
\end{document}